\documentclass[10pt,a4paper]{article}
\usepackage[latin1]{inputenc}
\usepackage{amsmath}
\usepackage{amsfonts}
\usepackage{amssymb}
\usepackage{graphicx}

\author{Joshua I. James\textsuperscript{1}, Yunsik Jake Jang\textsuperscript{2}\\
joshua@cybercrimetech.com, ccismem@gmail.com\\
\\
\textsuperscript{1}Digital Forensic Investigation Research Group\\
University College Dublin\\
Belfield, Dublin 4, IE\\
\\
\textsuperscript{2}International Cybercrime Research Center\\
Korean National Police University\\
Yongin-si, South Korea}
\title{Measuring digital crime investigation capacity to guide international crime prevention strategies}
\date{}
\begin{document}
\maketitle

\begin{abstract}
\noindent This work proposes a method for the measurement of a country's digital investigation capacity and saturation for the assessment of future capacity expansion. The focus is on external, or international, partners being a factor that could negatively affect the return on investment when attempting to expand investigation capacity nationally. This work concludes with the argument that when dealing with digital crime, target international partners should be a consideration in expansion, and could potentially be a bottleneck of investigation requests.
\end{abstract}

\section{Introduction}
Continuous development of information communication technology and large scale proliferation of digital devices is leading to an increase in victims of digital crime, as well as tools and evidence in criminal investigations [1]. Many crimes, even traditionally non-digital crimes such as murder, now normally have some sort of digital component [2,3]. A recent U.N. report also showed a global need for expansion of digital investigation capability [4], especially with cross-border technical assistance.

When expansion of investigation capacity is made, an organization should be able to quantify how this expansion affects the organization to ensure a maximized return on investment. Some works have previously examined investigation capacity in law enforcement [5], but have not specifically focused on digital investigation services or the interplay between national expansion and the effect on partner countries. When an organization has to interact with other organizations nationally and internationally -- especially in some sort of throughput or dependent relation -- they should also consider how their expansion will affect these other organizations, and how the other organizations may affect return on investment.

\subsection{Contribution and Structure}
This work examines the needs of digital investigation capacity expansion, and how such capacity expansion can be measured. This work argues that measurement of capacity and its effect on external partner organizations can lead to more strategic investment strategies to reduce global digital crime.

First, an overview of digital crime investigation capacity expansion in a global context is discussed. Section 3 proposes a method to model digital crime investigation capacity within a country. Next, the concept of investigation capacity saturation is discussed, and a method for choosing expansion investment based on strategic partner organization capacity saturation is given. Section 4 then gives final thoughts and areas for future work.

\section{Digital Crime Investigation Capacity Expansion in a Global Context}
Digital crime investigation units are oftentimes looking to expand investigation capacity. Expansion of investigation capacity could allow an organization to increase their scope of service, investigate all national case requests in a timely manner (reduce or eliminate a backlog), open more cases for international investigations, or even just allow an organization to better meet the needs of their own citizens. Expansion could come in the form of more funding for training, equipment, personnel, etc. [6,7]. For example, Irish digital forensic investigators estimated that up to 40\% of all exhibits receiving a full digital forensic analysis are determined to not be relevant to the case, and time and storage reductions could be made by slightly changing the investigation work flow [8].

Many organizations attempt to expand their investigation capacity when their budget allows. However, in the case of digital crime, many crimes have an international component. However, when one country needs assistance from another, the the investigation capacity of both countries should be considered. For example, if one country increases it's investigation capacity, this increase in capacity may result in more international requests. These requests will have a direct impact on the requested country's ability to handle other investigation requests. Consider Figure 1 and Figure 2. In Figure 1, if the current investigation capacity of Requestor 1 allows for 3 international requests, other countries may be able to handle these extra requests. However, after expansion of the investigation capacity of Requestor 1, international requests may now be doubled (Figure 2).

\begin{figure}
  \centering
  \includegraphics[width=0.9\textwidth]{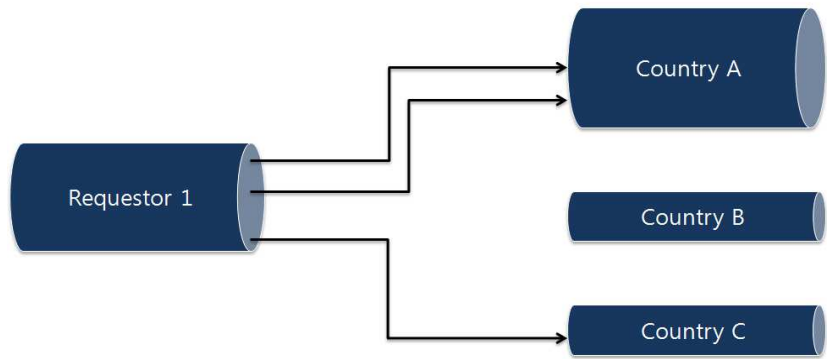}
  \caption{Requestor 1 making international requests with capacity before expansion}
\end{figure}

\begin{figure}
  \centering
  \includegraphics[width=0.9\textwidth]{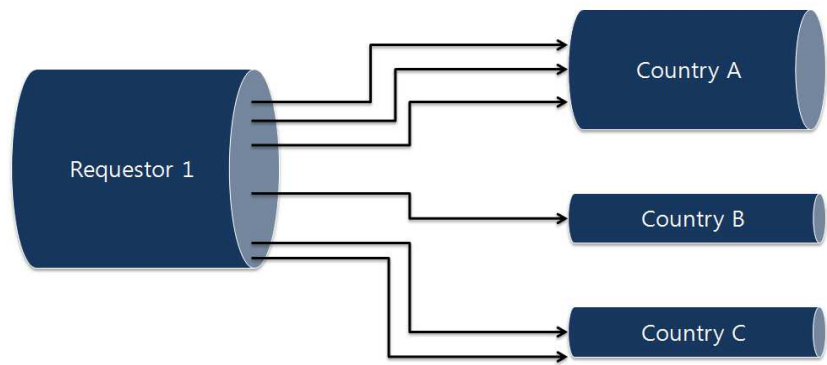}
  \caption{Requestor 1 making international requests with a national capacity after capacity expansion}
\end{figure}

While expansion may seem to be a benefit for Requestor 1, consider Figure 3, where Country C is receiving international requests. Country C's investigation capacity is significantly lower. So much so that Country C can only keep up with their national investigation requests. If Requestor 1 is now making more international investigation requests to Country C, Requestor 1 may have to wait longer for a response, meaning that their case throughput capacity is throttled by the investigation capacity of international partners.

\begin{figure}
  \centering
  \includegraphics[width=0.9\textwidth]{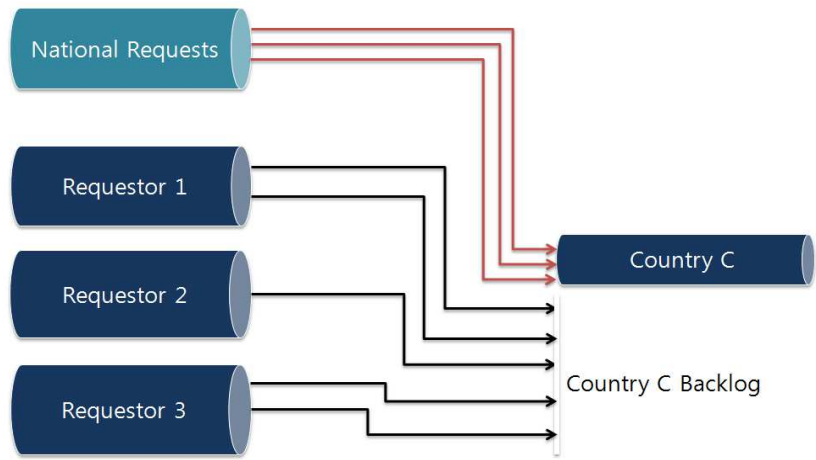}
  \caption{Investigations requested to Country C where the national requests take up all current capacity, and international requests are forced to a backlog}
\end{figure}

Countries considering expansion of investigation capacity should factor in the effect such a national-level expansion may have on other countries. And specifically determine at what point is there are reduced benefit to expanding investigation capacity.

\section{Modeling Digital Crime Investigation Capacity}
Digital crime investigation capacity can effectively be defined as \textit{the maximum amount of cases categorized as digital crimes that can be investigated by a group over a given period of time}.

A case backlog is defined as a \textit{queue of cases that are not being actively investigated, and are waiting to be started or finished by a group}. It may be common to have periods with a higher number of case requests than others. During peak periods, a backlog of cases may be created.

With these definitions, if total investigation requests over a given period of time are below investigation capacity then the cases will be started and cleared with no -- or minimal -- impact on a unit's backlog. If total investigation requests over a given period of time are above capacity, then the cases will be started and cleared in a longer period of time, and have a measurable impact on the case backlog.

Essentially, capacity in investigations cannot be thought of as applicable to each singular case. For example, if one investigator does not currently have work assigned, that `unused' capacity may not be able to be allocated to other ongoing tasks.

Capacity should be measured over time, and averaged for the group. This means that along with calculating the average number of cases completed per investigator, per year, the number of investigators also must be averaged per year. This allows units with high turn-over, or units with only part-time investigators assigned to still be able to attempt to measure capacity over the long term.

\subsection{An equation for investigation capacity}
In this work investigation capacity is defined as an equation over time where the cases completed at the investigator, unit or national level is divided over the average number of available investigators for the same time period.

\begin{itemize}
\item Let $T$ be the time-span of interest
\item Let $Ia$ be the time an investigator is available to work on cases
\end{itemize}

For $T: (cases closed)/(average investigators)\hspace{.2\textwidth}\dagger$\\
\noindent where:
\begin{itemize}
\item average investigators $= (Ia1/T)+(Ia2/T)+(Ia3/T)\dots +(Ian/T)$
\end{itemize}

This formula gives the average cases completed per available investigator. For example, if there were 4 cases closed over a 6 day period, one investigator was available full time, and one investigator was available part time (50\%), then the calculation would be as so: $4/((6/6)+((6/2)/6)) = 4/(1 + (3/6)) = 4/(1+0.5) = 4/1.5 = 2.7$

This means the throughput per investigator over this 6 day period is approximately 2.7 cases. Since there are essentially 1.5 investigators available, the overall potential throughput is $2.7 \cdot 1.5 = 4$. Throughput, however, is not necessarily the same as investigation capacity.

For units or countries that currently have a case backlog, the average number of cases completed per time-span, per investigator multiplied by the current number of available investigators is an indication of the current unit or national investigation capacity.

For units or countries that do not have a case backlog, the additional available investigation capacity may have to be estimated at least in two ways. Either by sampling maximum case throughput at a specific point in time where a backlog temporarily existed due to a surge in requests, or by using qualitative methods to ascertain at what capacity the investigators and their managers feel they performing at, or both.

For example, if the average cases completed annually in a country with 10 full-time investigators is 500, this country has no backlog, and each investigator estimates an average of 20\% of `down time', then the estimated national capacity would be 600 cases per year. $((500/10)+((500/10)\cdot 0.2)) \cdot 10 = 600$

With the consideration of downtime, the equation $\dagger$ should be updated as follows:
\\
\\
For $T: ((cases closed/average investigators)+\\((cases closed/average investigators) \cdot down time)) \cdot (average investigators)\hspace{.05\textwidth}\ddagger$
\\
\\
This formula will calculate, at least, an overall capacity estimate for the group over a period of time. Capacity measurements can be averaged for all groups/units in a country.

\subsection{Investigation capacity saturation}
Investigation capacity can be compared with investigation requests to determine the capacity saturation in a particular unit or country. Capacity saturation can be thought of as a country's ability to handle more case requests without the request being backlogged or jumping the queue. To calculate capacity saturation, the number of incoming case requests can be divided by the investigation capacity. Using this calculation, if capacity saturation is above 1 (100\%), then this indicates a backlog. The higher the capacity saturation, the more likely an organization will take a longer time to respond to a request, or potentially not respond at all.

If the investigation capacity of County A continually increases, and Country A increases the number of international requests made, the capacity of other countries may become over-saturated. At this saturation point, additional investment in Country A's investigation capacity could result in a reduction of throughput until other countries' capacity increases. In this case, it is more beneficial for Country A to invest in investigation capacity of countries to whom investigation requests are commonly made, but that have a lower investigation capacity.

\section{Conclusions}
Countries develop investigation capabilities and capacity at different rates, depending on budgets, focus, law, etc. This work gave an overview of digital crime investigation capacity expansion in a global context, and especially how expansion of investigation capacity at local organizations could have an effect internationally that may be negative. After, a method to model digital crime investigation capacity was proposed. Investigation capacity saturation was discussed, and a method for choosing expansion investment based on strategic partner organization capacity saturation was demonstrated.

\section*{References}
\begin{enumerate}
\item Casey, E. (2004). Digital evidence and computer crime : forensic science, computers and the Internet (2nd ed., pp. xviii, 690p.). Amsterdam, London: Elsevier Academic.
\item RTE. (2010). Text message evidence in murder trial. RTE News. Retrieved from http://www.rte.ie/news/2010/0423/drimnagh.html
\item Nguyen, L. (2012). Tori Stafford trial: Cellphone record shows gap during abduction, murder. Postmedia News. Retrieved from http://www.canada.com\\/life/Tori+Stafford+trial+Cellphone+record+shows+during+abduction\\+murder/6486178/story.html
\item (2013) ``Comprehensive Study on Cybercrime''. United Nations Office on Drugs and Crime. http://www.unodc.org/documents/commissions/\\CCPCJ\_session22/13-80699\_Ebook\_2013\_study\_CRP5.pdf
\item Hekim, H., Gul, S., \& Akcam, B. (2013). Police use of information technologies in criminal investigations. European Scientific Journal, 9(4), 221-240. Retrieved from http://eujournal.org/index.php/esj/article/view/778
\item Gogolin, G. (2010). The Digital Crime Tsunami. Digital Investigation, 7(1-2), 3-8. doi:10.1016/j.diin.2010.07.001
\item Casey, E., Ferraro, M., \& Nguyen, L. (2009). Investigation Delayed Is Justice Denied: Proposals for Expediting Forensic Examinations of Digital Evidence. Journal of forensic sciences, 54(6), 1353-1364. doi:10.1111/j.1556-4029.2009.01150.x
\item James, J. I., \& Gladyshev, P. (2013). A survey of digital forensic investigator decision processes and measurement of decisions based on enhanced preview. Digital Investigation, 1-10. doi:10.1016/j.diin.2013.04.005
\end{enumerate}
\end{document}